# Numerical and Experimental Investigation of Static Wetting Morphologies of Aqueous Drops on Lubricated Slippery Surfaces Using a Quasi-Static Approach

*Shivam Gupta, Bidisha Bhatt, Meenaxi Sharma, and Krishnacharya Khare\**

*Department of Physics, Indian Institute of Technology Kanpur, Kanpur – 208016, India*

\*email: kcharya@iitk.ac.in

**Abstract**

*Hypothesis:* Due to the slow dynamics of the wetting ridge, it is challenging to predict the wetting morphology of liquid drops on thin lubricant coated surfaces. It is hypothesized that when a drop sinks on a lubricated surface, quasi-static wetting morphology can be numerically computed only from the knowledge of interfacial energies, lubricant thickness, and drop volume.

*Simulations and experiments:* We used Surface Evolver software for the numerical computation of the interface profiles for a four-phase system. For the experiments, we used drops of 80 wt% formamide on silicone oil coated substrates with varying lubricant thickness, substrate wettability and drop volume. Optical images of drops were used to compare the experimental interfacial profiles and apparent contact angles with the numerically computed ones.

*Findings:* We found good agreement between the experiments and the simulations for the interfacial profiles and apparent contact angles as a function of various systems parameters except for very thin lubricating films. Apparent contact angles varied non-linearly as a function of substrate wettability and lubricant thickness, however, were found constant with the drop volume.

## 1. Introduction

For more than a decade, lubricant-infused slippery surfaces inspired by *Nepenthes'* pitcher plants have gained a lot of attention [1-3]. The surfaces comprise a solid substrate with micro/nano textures infused with a highly spreading lubricating fluid (oil) that is immiscible with other test liquids (aqueous). When drops of a test liquid come into contact with such surfaces, different configurations of the lubricating oil film and, subsequently the drops have been theoretically predicted and experimentally verified [4-7]. The configurations of the lubricating fluid and test liquid drops depend upon various interfacial energies and intermolecular interactions. If the sum of surface energies of the oil-drop interface ($\gamma_{OD}$) and the oil-air interface ($\gamma_{OA}$) is energetically more favorable than the surface energy of the drop-air interface ($\gamma_{DA}$), i.e., if $\gamma_{OD} + \gamma_{OA} < \gamma_{DA}$, then the test liquid drops get engulfed with nanometer thin layer of the lubricating fluid [4, 5]. This phenomenon is commonly referred to as cloaking. The drop-air surface energy per unit area, $\gamma_{DA}$ then changes to effective surface energy per unit area, $\gamma_{DA,eff} = \gamma_{OD} + \gamma_{OA} + \frac{A_H}{12\pi t_{cloak}^2}$ [8]. Considering the typical values of the Hamaker constant, $A_H \sim 10^{-21}$J [6, 9] and the cloaking thickness, $t_{cloak} \sim$ tens of nm [4, 10, 11], the term $\frac{A_H}{12\pi t_{cloak}^2}$ can be neglected and thus to first approximation $\gamma_{DA,eff}$ for cloaked drops can be approximated as $\gamma_{OD} + \gamma_{OA}$. The approximation thus helps in conveniently studying and/or simulating the system in continuum regime.

Besides engulfing the drop, the thin oil film under the drop has also been found to be in three states: L1, L2, and L3. Depending upon the interfacial energies and the intermolecular interactions, the oil is either completely stable (L1) or the drop directly contacts the substrate by either dewetting the oil into multiple drops (L2) or completely squeezing out the oil (L3) [6]. The oil displays state L1 when the Hamaker constant, $A_H > 0$ and the spreading coefficient, $S = \gamma_{OA}\cos\theta_{OS} - \gamma_{DA}\cos\theta_{DS} - \gamma_{OD} > 0$. Here O, D, and A corresponds to the oil, drop, and air phase, $\gamma_{ij}$ is the surface energy per unit area of the *i-j* interface, respectively, and $\theta_{DS}$ and $\theta_{OS}$ is the contact angle of the drop and the oil on the substrate in ambient air, respectively. The expression for $S$ can also be rewritten in terms of the drop's contact angle on the substrate in ambient oil, denoted as $\theta_{DS}^O$. Theoretically, $\theta_{DS}^O$ is calculated using the Bartell-Osterhof equation [12],

$$\gamma_{OD} \cos \theta_{DS}^O = \gamma_{DA} \cos \theta_{DS} - \gamma_{OA} \cos \theta_{OS} \tag{1}$$

which on inserting in the expression for spreading coefficient gives $S = -\gamma_{OD}(1 + \cos \theta_{DS}^O)$. Thus $S > 0$ only if $\theta_{DS}^O = 180°$. For the state L1, the drop is referred to be in the floating state since there is always a thin film of oil that prevents the drop from directly contacting the substrate. This leads to weak adhesion between the drop and the substrate. Making use of this understanding, slippery surfaces have been shown to be effective for various applications such as efficient heat transfer [13, 14], fog harvesting [15-18], anti-icing [14, 19-21], anti-fouling [22-24] and self-cleaning applications [7, 25, 26], to name a few. However, if one of the conditions is not satisfied, then the oil displays either state L2 or L3, and the drop is thus referred to be in the sinking state. Recently it has been shown that such sinking systems can be used for controlled particle deposition and, therefore, can be used for self-assembly and surface coating applications [27]. Hence, not only floating but sinking systems are also intriguing from a scientific perspective.

Thin lubricating fluid coated surfaces can be further categorized based on the size of the wetting ridge. For systems where the radius of curvature of the wetting ridge is smaller (/bigger) than the capillary length of the system, the system is referred to be in the starved (/excess) lubricant regime [28]. Due to the slow dynamics of the wetting ridge, within practical times, most of the systems show a starved lubricant regime. Although a lot of studies on slippery surfaces have been reported for the starved regime, however, very few studies have examined the effect of various system parameters on interfacial profiles and apparent contact angles. Recently, Semprebon *et al.,* using Surface Evolver (SE), studied the effect of relative pressure between the drop-air and the oil-air interface on the drop shape and contact angle hysteresis [29, 30]. The oil-air interfacial pressure was set as the control parameter for the study. The prediction of the maximum value of the apparent contact angle for negligible wetting ridge, $\theta_{app}^S$ is given as,

$$\cos\theta_{\text{app}}^S = (\gamma_{OD}\cos\theta_{DS}^O + \gamma_{OA}\cos\theta_{OS})/\gamma_{DA} \qquad (2)$$

Besides predicting the maximum apparent angle, the study can't be directly applied to predicting various interfaces since, in the experiments, the oil volume instead of pressure can be controlled. Moreover, the study is only limited to uncloaked systems. Another theoretical study on finding the profile of the drop-oil interface from the knowledge of Neumann's point and the curvature of the drop-air and the oil-air interface has been reported by Gunjan *et al.* [8, 31]. For the prediction, however, the theory relies on the experiments to be performed first. Thus, there is no theory/simulation to date that can beforehand predict the interfacial profiles of the oil-air, drop-air, and oil-drop interfaces as a function of various system parameters: lubricant thickness, substrate wettability, and drop volume.

Compared to the excess lubrication systems, the beforehand prediction of the interfacial profiles is challenging for starved systems mainly because of two reasons. First, due to the viscous dissipation in the oil, not all of the available oil flows to the ridge within practical times. Second, because of the capillary suction, the minima of the ridge, $h_{\min}$ drops below the coated lubricant thickness [32]. Recently, the theoretical estimation of the time scales and height $h_{\min}$ for such two-dimensional problems has been studied by Dai *et al.* [33]. Accordingly, for the time of observation, $\tau$, ranging between $\tau^* \sim \mu t/\gamma$ and $\tau_0 \sim (\alpha \mathcal{V}_\gamma)^{-2/3} \alpha^4 \mu t/\gamma$ i.e. if $\tau^* \ll \tau \ll \tau_0$, only the oil in the vicinity of the drop would add to the volume of the wetting ridge and the height $h_{\min}$ in this early-intermediate time is given by $(\mathcal{V}_\gamma/\alpha)^{1/3} t$. Here $\alpha$ represents the aspect ratio of the film and $\mathcal{V}_\gamma$ is the measure of the relative importance of Van der Waals forces and the capillary pressure in the lubricating film, with their typical values ranging from $25 - 250$ and $10^{-7} - 10^{-3}$ respectively. Although the theory is for two-dimensional systems, we expect it to give a fair estimate of the time scale and height of the three-dimensional systems.

Using the numerical simulations, we have worked on a method for the beforehand prediction of the quasi-static interfacial profiles and the effect of various system parameters on apparent contact angles after a drop sinks on a starved flat lubricated surface with drop sinking time, $\tau_S \sim$ early-intermediate times. We define $\tau_S$ as the time taken by the drop to reach a quasi-static contact angle after dewetting the oil.

## 2. Methods and materials

### 2.1. Physical model

For a drop on a flat thin lubricating fluid coated slippery surface, the total free energy of the system can be written as,

$$E = \sum_{i \neq j} \iint_{A_{ij}} \gamma_{ij} dA + \sum_{\alpha} \iint_{A_{\alpha s}} \gamma_{\alpha s} dA + \sum_{\beta} \iiint_{\beta} \rho_\beta gz \, dV \qquad (3)$$

where the first term corresponds to the surface energy of the fluid-fluid interfaces, with $i, j$ representing the oil (O), drop (D), and air (A) phases and $A_{ij}$ corresponds to the surface area of the $i$-$j$ interface. The second term represents the surface energy of the fluid-substrate interface. For a sinking system, $\alpha$ corresponds to the oil and drop phase since both the fluid phases contact the solid substrate. For a floating system, however, $\alpha$ corresponds to the oil phase only since the drop and the substrate never come into direct contact. The third term is the gravitational potential energy of $\beta$ body, where $\beta$ corresponds to the oil or drop phase with the density $\rho_\beta$ under the gravitational acceleration, g. The term is significant only when the typical length scales of the drop (the drop size) and the oil (size of the wetting ridge) are larger than the respective capillary lengths.

The effect of substrate wettability on the drop shape can be better understood by writing the interfacial energy of the drop-substrate and oil-substrate interfaces in terms of their respective contact angles using the Young's equation. If the contact angle of the drop and the oil with the substrate in the air is $\theta_{DS}$ and $\theta_{OS}$, respectively, then Eqn. (3) can be rewritten as,

$$E = \sum_{i \neq j} \iint_{A_{ij}} \gamma_{ij} dA - \gamma_{DA} A_{DS} \cos\theta_{DS} - \gamma_{OA} A_{OS} \cos\theta_{OS} + \sum_{\beta} \iiint_{\beta} \rho_\beta gz \, dV + C \qquad (4)$$

where the constant $C = \gamma_{SA}(A_{DS} + A_{OS})$. Thus, the system reaches the minimum energy state by minimizing the energy $E - C$.

To quantify the static wetting morphology of drops, two different apparent contact angles, $\theta'_{app}$ and $\theta_{app}$ are used, where $\theta'_{app}$ is the angle made with the substrate when the drop's curvature is extrapolated and $\theta_{app}$ is the angle at the three-phase contact line (TPCL), as shown in Fig. (1). In the case of cloaked drops, the TPCL is replaced with the equivalent TPCL, which is defined as the contact point of the wetting ridge and the cloaked oil film. At the equivalent TPCL, the Neumann's angles [34, 35]: $\theta_D, \theta_A$ and $\theta_L$, as shown in Fig. (1), also get replaced with the equivalent Neumann's angles.

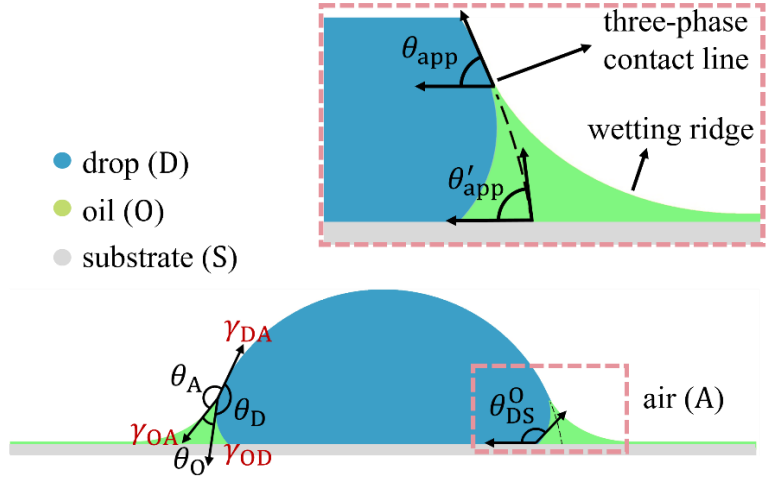

**Figure 1:** Schematic of a drop on a thin lubricating fluid coated surface illustrating the Neumann's (or equivalent Neumann's) angles, $\theta_D, \theta_A$ and $\theta_L$ at the three-phase contact line (TPCL) (or equivalent TPCL) and the system parameters, substrate wettability ($\theta^O_{DS}$) and surface energies $\gamma_{ij}$. The zoomed-in image shows the two different apparent contact angles, $\theta'_{app}$ and $\theta_{app}$ at the substrate and the TPCL, respectively.

*2.2. Numerical method*

To numerically model the system, we used Surface Evolver (SE) software developed by Brakke [36], which is a public domain interactive software based on the finite element method. The initial capillary surfaces in SE are defined in a text file by assigning their vertices, edges, faces, surface energies, and appropriate constraints. The program then minimizes the total energy of the system using the gradient descent method. We started by defining the oil and the drop as two separate bodies in SE with their respective mass densities. To visually differentiate the interfaces, light green, light blue, and light grey color codes were used, representing the oil-air, drop-air, and drop-oil interfaces, respectively. The initial surface of a drop was defined as a cuboid. A contact angle in SE cannot be defined unless the surfaces are in direct contact, hence the initial interface in SE was defined with the drop in contact with the substrate and surrounded by the oil, as shown in Fig. 2 (a). We assume that this atypical initial condition wouldn't affect the final drop shape. The bottom facets of the drop and the oil touching the substrate were removed, and the edges enclosing the facets were given the energy integrand, as explained in the SE manual [37].

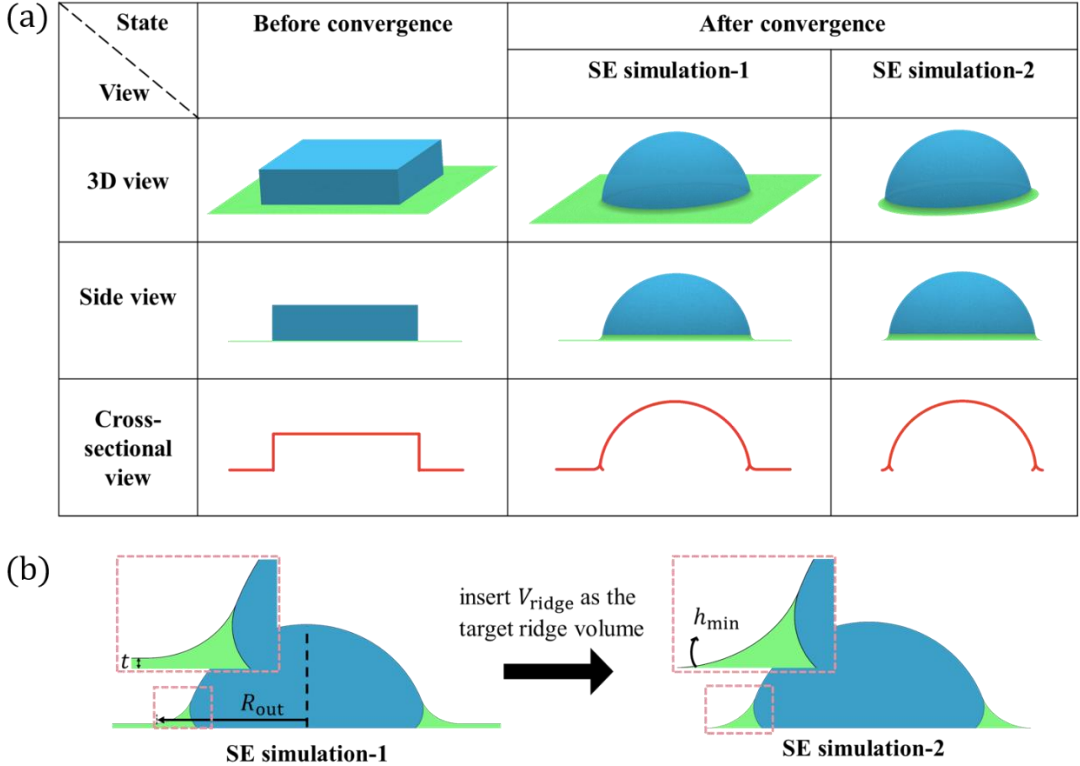

**Figure 2:** (a) Before and after convergence of Surface Evolver simulation of a drop on a slippery surface with various views: 3D, side, and cross-sectional views. The plane perpendicular to the substrate and passing through the center of the drop is used for the cross-sectional view. (b) Schematics illustrating the method for estimating the wetting ridge volume. SE simulation-1 (minimum height of oil-air interface equals coated lubricant thickness) is used to estimate the wetting ridge volume ($= \pi R_{out}^2 t$). The estimated ridge volume is then used as the target ridge volume for simulation-2 (oil-air interface touches the substrate). The insets show the zoomed-in image of the wetting ridge for the respective simulations.

This allows the inward hassle-free movement of edges on substrates. The volume of the drop, $V_D$, was set as a global constraint. For systems that satisfy the cloaking condition, we set drop-air surface energy as $\gamma_{DA,eff}$. For the rest of the facets, we used the corresponding surface energies. Although the Hamaker constant, $A_H$, plays a vital role in determining the morphology of the oil underneath the drop, however, we don't expect it to play a significant role in determining the quasi-static apparent contact angles and the interface profiles. Since we are interested in the drop's wetting morphology instead of the oil's morphology underneath the drop, we did not incorporate the long range interaction in our simulation.

As stated earlier, estimating the ridge volume is not straightforward due to the limitations of the starved lubricant system. Thus for the prediction, one needs to know the approximate target ridge volume. Using SE, we developed a quasi-static procedure to estimate wetting ridge volume for a starved flat lubricant system that satisfy the sinking criteria and has sinking time ($\tau_S$) similar to early intermediate times. To estimate the ridge volume, $V_{\text{ridge}}$, we carried out two simulations, as shown in Fig. (2). In SE simulation-1, we fix the outer edges of the oil and constrain the minima of the ridge to be equal to the thickness of the lubricant film (*t*) and also conserve the coated lubricant volume. The volume of the ridge will then be equal to $\pi R_{\text{out}}^2 t$, where $R_{\text{out}}$ is the horizontal distance from the center of the drop to the point where the slope of the wetting ridge becomes zero, as shown in Fig. 2 (b) (see S1, Supplementary Material). Next, for simulating the system of interest, we lift the constraint on outer edges and allow the vertical position of the minima to be equal to $h_{\text{min}}$. We assume that experimentally, the vertical shift in the position of the minima is not significant enough to change the wetting ridge volume. With this assumption, the volume then found from the previous part acts as the target ridge volume for the current simulation, SE simulation-2, as shown in Fig. 2(b). Although the procedure for estimating the ridge volume should ideally be only for the state L3 (where oil is completely squeezed out under the drop), however, for the state L2 (where few dewetted oil droplets are left under the drop), we can still estimate $\pi R_{\text{out}}^2 t$ as the approximate wetting ridge volume if the ridge volume is significantly larger than the total volume of dewetted oil droplets.

*2.3. Experiments*

In experiments, aqueous solution of 80 wt% formamide (Loba Chemie, India) in DI water was used as the liquid for the top drops. The atypical aqueous solution was chosen due to its low viscosity and nearly hygroscopically stable nature in the lab condition (see S2, Supplementary Material). Silicone oil

(Gelest Inc., USA) with density 969 kg/m$^3$, kinematic viscosity 350 cSt, and surface tension 21.1 mN/m, was used as the lubricating fluid. Using the pendant drop method, the interfacial tensions of the drop-air and the drop-oil interfaces were found as 61.1 (1.2) mN/m and 26.0 (0.9) mN/m, respectively (see S3, Supplementary Material). Silicon (Si) surfaces (Prime grade, UniversityWafer Inc., USA) were used as the solid substrates. To clean the surface of Si substrates, they were put in an ultrasonicator bath for 10 minutes each in ethanol, acetone and toluene followed by O$_2$ plasma (Harrick Plasma, USA) for few minutes. For preparing surfaces with different wettability, the cleaned Si surfaces were first made hydrophobic by grafting them with octadecyl trichlorosilane (OTS) (Sigma–Aldrich, USA) using standard procedure and were then placed in high O$_2$ plasma for different times (1s-5s)[38, 39]. To change the thickness of the lubricant, the rotation speed of the spin coater was varied. The thickness of the spin-coated lubricathing films were measured using scanning optical reflectometer (Filmetrics F20-EXR, USA) (see S4, Supplementary Material). To dispense the drop of controlled volume, $V_D$, the automatic dispense unit of the optical contact angle goniometer (OCA-35, DataPhysics, Germany) was used. Since the drop movement leads to the increase in the volume of wetting ridge [10], the needle was kept attached to the drop until the sinking of the drop. The apparent contact angles, $\theta_{app}$ and $\theta'_{app}$, of aqueous drops on lubricant coated surfaces were also measured from the contact angle goniometer. All contact angle measurements were carried out for three different drops at three different places to estimate the error.

3. Results and discussion

*3.1. System properties*

Since the present combination of drop (aqueous solution of 80 wt% formamide) and lubricating fluid (silicone oil) satisfies the cloaking condition, drop-air interfacial energy is set as $\gamma_{DA,eff}$ in SE simulations [8]. Since hydrophilic silicon are used as substrates, thin lubricating films underneath aqueous drops become unstable (dewet), leading to the sinking of the drops. Experimentally, the sinking time is observed to be 5-10 minutes, which lies well within $\tau^*$ (~ ms) and $\tau_0$ (~ 100 days). This means that the ridge volume can be estimated as the sum of the volumes of oil squeezed out by the drop and the oil just in the vicinity of the drop. From the values of $\mathcal{V}_\gamma$ (~ $10^{-7}$) and $\alpha$ (~ 62) corresponding to the present system, the minimum thickness of the oil-air interface can be calculated as $h_{min}$ ~ $10^{-8}$ m, which can be neglected compared to the thickness of lubricating films. Hence, $h_{min} = 0$ is used in SE simulation-2. For all the values of substrate wettabilities used in the experiments, dewetted oil droplets were always seen under the top drops, which means that the system corresponds to the L2 state. Since the present experimental system satisfies all the required conditions to efficiently run SE simulations, it can be used to predict the wetting morphologies of drops on slippery surfaces.

*3.2. Finding apparent angles for simulation*

Upon providing the contact angles, $\theta_{DS}$ and $\theta_{OS}$ to SE, the contact angle of a sessile drop surrounded by oil, $\theta_{DS}^O$, is calculated by SE using the Bartell-Osterhof equation. However, one has to be careful while comparing the numerically calculated $\theta_{DS}^O$ with the corresponding experiments since previous few studies have reported significant disagreement [40, 41].

| $\theta_{DS}$ (exp) | $\theta_{DS}^O$ (calc) | $\theta_{DS,e}^O$ (exp) | $\theta_{DS}$ (calc) |
|---|---|---|---|
| 75.2 (1.1) ° | 102.1 ° | 110.1 (3.1) ° | 78.6 ° |
| 82.3 (3.6) ° | 119.6 ° | 119.2 (2.6) ° | 82.1 ° |
| 87.5 (0.8) ° | 134.9° | 143.1 (2.4) ° | 89.8 ° |
| 93.5 (1.0) ° | 161.8 ° | 163.9 (3.0) ° | 93.7 ° |

**Table 1:** Comparison of the experimental ($\theta_{DS,e}^O$) and calculated ($\theta_{DS}^O$) apparent contact angles for aqueouse drop in silicone oil medium on substrates with different wettability ($\theta_{DS}$). $\theta_{DS}^O$(calc) and $\theta_{DS}$(calc) are calculated from the Bartell-Osterhof equation (Eq. 1).

The reason for such deviation is the presence of thin oil film between the drop and the substrate, which can to be present even for the sinking system where the underneath lubricant films are not stable[42].

Thus, one needs to first find the angle $\theta_{DS}^O$ experimentally (denoted as $\theta_{DS,e}^O$) as shown in Fig. (3), and subsequently find $\theta_{DS}$ (referred to as $\theta_{DS,app}$) from the Bartell-Osterhof equation and then use it in SE to calculate the correct angle, $\theta_{DS}^O$, as listed in table (1).

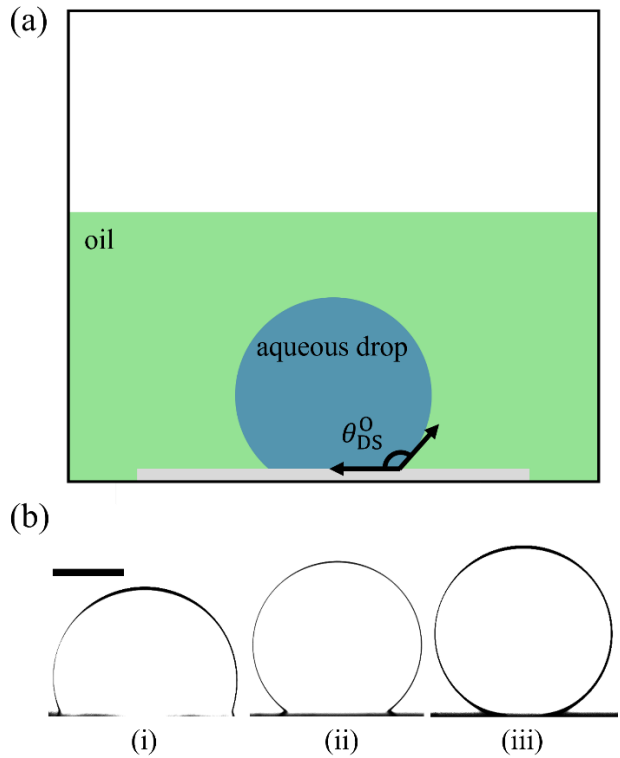

**Figure 3:** (a) Schematic of a drop on a solid substrate in ambient oil showing a contact angle $\theta_{DS}^O$. (b) Optical contact angle images of aqueous (80 wt% formamide) drops on different wettability substrates in ambient silicone oil. The observed values of the contact angle, $\theta_{DS,e}^O$ for (i), (ii), and (iii) are 110.1 (3.1)°, 143.1 (2.4)°, and 163.9 (3.0)°, respectively. The scale bar corresponds to 500 μm.

*3.3. Effect of the thickness of lubricant film*

To study the effect of various system parameters, such as the thickness of lubricant films, the wettability of the substrates, and the volume of drops, on the static wetting morphologies, the apparent contact angles and interface profiles were measured experimentally and also computed numerically using SE simulations. Due to the complete wetting of the silicone oil on the substrates, we put $\theta_{OS} = 0°$ in all the SE simulations. To study the effect of the thickness of lubricathing films, the substrate wettability and the drop volume, were kept fixed to 143.1 (2.4)° and 2.0 (0.2) μl, respectively. Figure 4 summarizes the effect of the thickness of lubricating films on the static wetting morphologies of drops. The apparent contact angle at the substrate ($\theta'_{app}$) does not change much with the lubricant film thickness, however the apparent contact angle at the TPCL ($\theta_{app}$) decreases montonically. Figure 4 (b) shows the interface profiles of aqueous drops obtained from SE simulations, and it is clear that the maximum height of the wetting ridge or the TPCL moves vertically upwards with the increase in the lubricant film thickness. As a result, $\theta_{app}$ decreases but $\theta'_{app}$ does not change much. The reason for such behavior is the increase in the relative Laplace pressures at the drop-air and the oil-air interfaces with the increasing lubricant thickness. This is consistent with the results of Semprebon et al. [29].

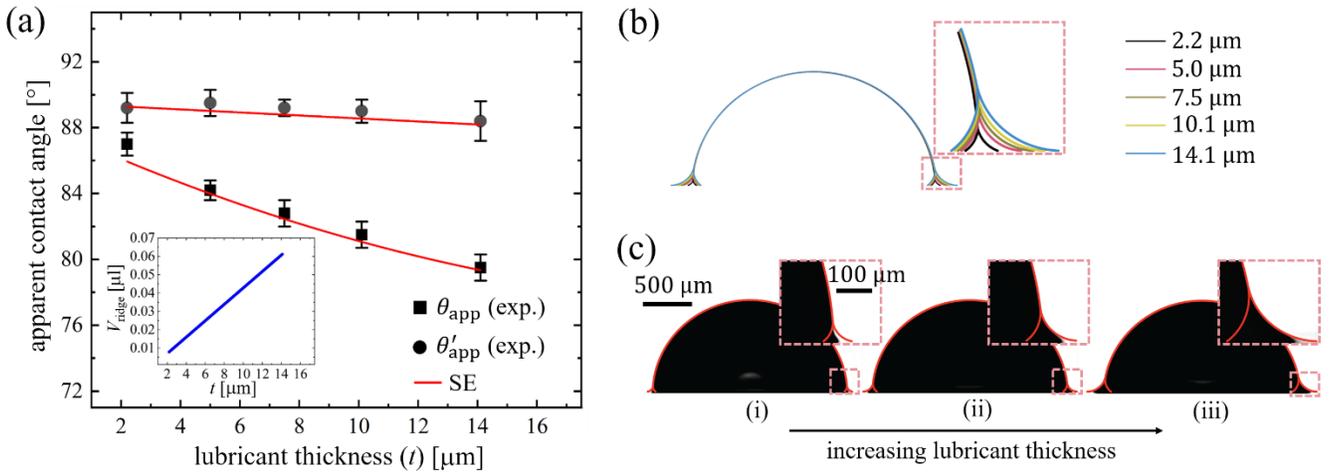

**Figure 4:** Effect of the thickness ($t$) of lubricating films on the apparent contact angles and interface profiles of aqueous drops on lubricated surfaces for fixed substrate wettability and drop volume. (a) The graph shows the variation of $\theta_{app}$ and $\theta'_{app}$ with the thickness of lubricating films of silicone oil. The black data points represent experimental values, and the solid red lines correspond to the SE simulations. It is clear that $\theta'_{app}$ remains almost constant, however $\theta_{app}$ decreases with $t$ since the TPCL moves vertically upwards with increasing $t$, as shown by the numerically computed interface profiles in (b). Inset in (a) shows the variation of the estimated volume of the wetting ridge with $t$. (c) Comparing the optical contact angle images of aqueous drops with the corresponding interface profiles obtained by SE simulations for three different thicknesses of the lubricating films, (i) 2.2 (0.5) µm, (ii) 7.5 (0.5) µm and (iii) 14.1 (0.5) µm, showing a good agreement.

The maximum apparent contact angles angle found from the experiments and simulations are 89.2 (0.9)° and 89.2°, respectively, which is in good agreement with the value 89.6° found from Eq. (2) on replacing $\gamma_{DA}$ with $\gamma_{DA,eff}$. Figure 4 (c) compares the optical contact angle images of aqueous drops with the corresponding interface profiles obtained from SE simulations. From the insets in Fig. 4 (c), it is observed that although the drop-air interface profile matches the simulations, the estimation of the wetting ridge profile, however, is not in good agreement with the experiments for very small lubricant thickness, corresponding to the negligible wetting ridge. This is because the volume of the dewetted drops starts to become more and more significant as the lubricant thickness is decreased which leads to an error in the estimation of wetting ridge volume from the method. Drop profile agreement becomes better for a thicker lubricant system. In addition, we also don't expect such discrepancy for systems where the oil is completely displaced from underneath the drops (the state L3). We also calculated the volume of wetting ridge from SE simulations and found that it also increases linearly with the film thickness, as shown in the inset of Fig. (4)(a).

*3.4. Effect of the substrate wettability*

To study the effect of the substrate wettability, the thickness of the lubricating film and the drop volume were kept constant at 5.0 (0.5) µm and 2.0 (0.2) µl, respectively. Figure 5(a) shows that, both $\theta_{app}$ and $\theta'_{app}$, increases non-linearly with increasing $\theta_{DS}^O$. In addition, the estimated volume of wetting ridge also decreases nonlinearly, as shown in the inset Fig. 5(a). However, both the apparent contact angles and the wetting ridge volume saturate around $\theta_{DS}^O = 180°$, which is referred to as the floating point angle. Such variation is expected close to the floating point angle since at this angle, due to the presence of thin oil film between the drop and the substrate, the drop morphology no longer gets affected with the substrate wettability. This result is a correction to the results of Sharma et al. [43] for apparent angles for a similar system. Therefore, the variation of apparent angles saturates close to the floating-point angle, $\theta_{DS}^O = 180°$ (or equivalently $\theta_{DS} = 104.5°$ for their case) and not at the boundary of hydrophilic and hydrophobic surfaces ($\theta_{DS} = 90°$) as reported by them. Figure (5)(b) shows the interface profiles of aqueous drops obtained from SE simulations as a function of increasing substrate wettability, which shows that the maximum ridge height does not change significantly with the substrate wettability, for fixed lubricant thickness and drop volume.

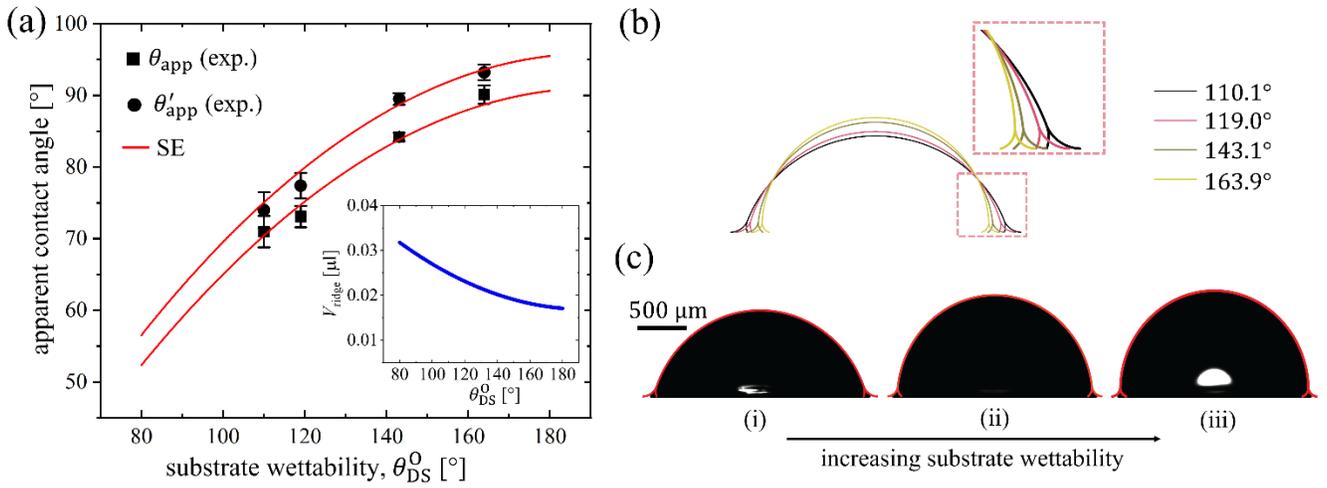

**Figure 5:** Variation of apparent contact angles and interfacial profiles with substrate wettability for fixed drop volume and lubricant thickness. The graph in (a) shows the behavior from the numerical simulations and the experiments, with the inset showing the variation of the estimated volume of the wetting ridge. (b) Shows the simulated interfacial profiles. (c) Shows the overlap of simulation and optical images with $\theta_{DS}^{O} =$ 110.1 (3.1)°, 143.1 (2.4) ° and 163.9 (3.0) ° for (i), (ii) and (iii).

Figure (5)(c) shows the overlap of optical contact angle images of aqueous drops on lubricated substrates with different substrate wettability, (i) 110.1 (3.1)°, (ii) 143.1 (2.4)° and (iii) 163.9 (3.0)°, with the corresponding interface profiles obtained from SE simulations. The good agreement between them confirms the validity of the SE simulations.

*3.5. Effect of the drop volume*

To study the effect of drop volume on wetting morphology, the lubricant film thickness and the substrate wettability were fixed at 5.0 (0.5) μm and 143.1 (2.4)°, respectively. As shown in Fig. (6)(a), no change in the apparent angles is observed, however the estimated volume of the wetting ridge increases linearly with the drop volume. The behavior is similar to the non-lubricated surfaces, where the volume of the drop only affects the drop's shape (drops get deformed by gravity) and not the contact angle of the drop. Figure 6(b) shows the optical images of aqueous drops of different volume, (i) 2 μl, (ii) 6 μl, and (iii) 10 μl, together with the interface profiles obtained by SE simulation of the corresponding system.

Inset in Fig. 6(a) shows that the wetting ridge volume increases with the drop volume, however the maximum height of the wetting ridge does not change significantly.

In addition to the 80 wt% formamide solution, static wetting morphologies were also investigated for two more systems, viz. ethylene glycol and water with silicone oil as lubricating fluid to generalize our method (see S5, Supplementary Material). Therefore, using the quasi-static approach, numerical simulation of aqueous drops on thin lubricating fluid coated surfaces using Surface Evolver provides accurate and reliable static wetting morphologies, as also confirmed using the corresponding experiments.

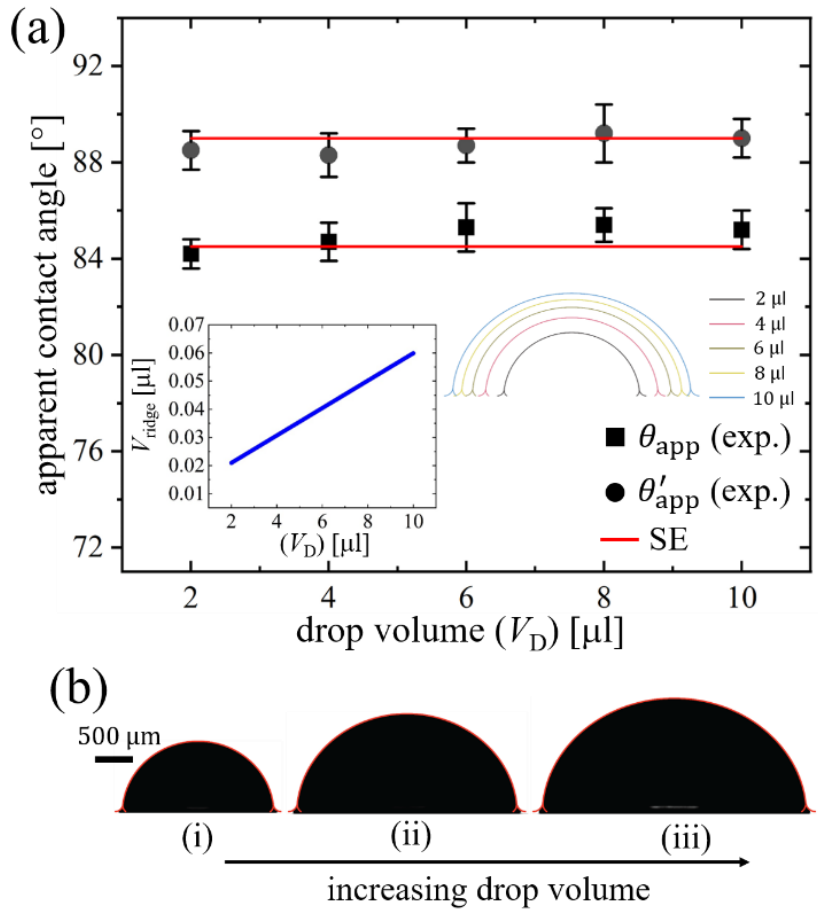

**Figure 6:** Variation of apparent contact angles and interfacial profiles with drop volume for fixed substrate wettability and lubricant thickness. The graph in (a) shows the behavior found from both numerical simulations and the experiments, with the inset showing the variation of the estimated volume of the wetting ridge and the simulated interfacial profiles. (b) Shows the overlap of simulations and optical images of aqueous drops with $V_D = $ 2 (0.2) µl, 6 (0.2)µl and 10 (0.2) µl for (i), (ii) and (iii).

## 4. Conclusion

It is often required to know the static wetting morphology of drops on thin lubricating fluid coated slippery surfaces, which has been investigated numerically and experimentally in this study. Using a quasi-static approach, apparent contact angles of sessile drops and various interface profiles are computed numerically using Surface Evolver software for sinking lubricated system, with the sinking time similar to the early intermediate times, under starved lubricant condition. We found that the interface profiles can be computed numerically from the knowledge of the system parameters: surface energies, lubricant thickness, and drop volume. Compared to the previous studies, which rely on the beforehand experimental information to compute the interface profiles [8, 29-31], our study adopts a quasi-static approach to compute the static wetting morphologies. With only the prior information of the various interfacial energies and system parameters, the current method can be efficiently used to predict apparent contact angles and various interface profiles of the system. From a primary simulation, the wetting ridge volume is computed for a given system, which is then used in subsequent simulation as the target ridge volume to predict the wetting morphologies. Experiments with corresponding system parameters are in good agreement with the numerical results, which confirms the validity of the quasi-static approach of numerical simulations.

The apparent angles at the substrate ($\theta_{\text{app}}$) and at the TPCL ($\theta'_{\text{app}}$), were found to vary differently with increasing lubricant thickness, however, with increasing substrate wettability and drop volume, both angles varied in a similar way. The results are consistent with the study by Semprebon et al.[29] A good agreement was found between the experiments and the corresponding simulations for increasing lubricant thickness except for very thin lubricant films. The variation of the apparent angles with the substrate wettability was observed to saturate around the floating-point angle both in the experiments and in the simulations. This means that the saturation value is controlled by the system parameters instead of the conventional hydrophilic-hydrophobic boundary as reported earlier [7, 43, 44].

Although there is a good agreement between the numerical simulations and the experiments for quasi-static apparent contact angles and quasi-static interface profiles, a complete dynamic prediction of the profiles and thus the variation of apparent contact angles with system parameters (for both sinking and floating slippery systems) still remians an open research problem.

5. Acknowledgement

Authors acknowledges the funding support from SERB, New Delhi (Project no. CRG/2019/000915) and DST, New Delhi, through its Unit of Excellence on Soft Nanofabrication at IIT Kanpur. SG acknowledges funding support from Prime Minister's Research Fellows (PMRF) Scheme of the Government of India. Authors acknowledge the useful discussion with Prof. K. Brakke.

**Numerical and Experimental Investigation of Static Wetting Morphologies of Aqueous Drops on Lubricated Slippery Surfaces Using a Quasi-Static Approach**

*Shivam Gupta, Bidisha Bhatt, Meenaxi Sharma, and Krishnacharya Khare\**

*Department of Physics, Indian Institute of Technology Kanpur, Kanpur – 208016, India*

**S1: Volume estimation of the target ridge volume from SE simulation-1:**

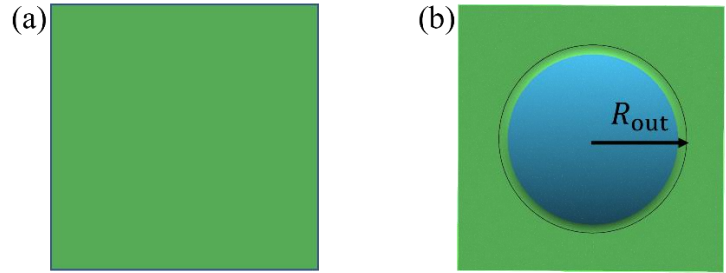

**Figure S1:** Top view of a lubricant-coated slippery surface (a) before and (b) after drop deposition.

Figure S1 shows the top view of a drop deposited on a thin lubricating fluid coated surface with lubricant thickness, $t$, lubricant surface area, $A_{\text{lubricant}}$, and volume, $V_i$. If the sinking time of the drop, $\tau_S$, is similar to early intermediate times, the volume of the wetting ridge, $V_{\text{ridge}}$, can be derived by conserving the lubricant volume as follows,

$$V_i = V_{\text{ridge}} + V_{\text{non-ridge}} \quad (1)$$

Using $V_i = A_{\text{lubricant}}\, t$, $V_{\text{non-ridge}} = A_{\text{non-ridge}}\, t$ and $A_{\text{non-ridge}} = A_{\text{lubricant}} - \pi R_{\text{out}}^2$, we get

$$V_{\text{ridge}} = \pi R_{\text{out}}^2\, t \quad (2)$$

where $A_{\text{non-ridge}}$ and $V_{\text{non-ridge}}$ is the surface area and volume, respectively, of the lubricant from the region far away from the drop that has not flown to the wetting ridge.

## S2: Variation of the drop volume with time:

Environmental exposure to an aqueous drop (80 wt% formamide) on a dry hydrophobic surface under the lab condition (19.8 °C and 55% RH) results in only 10 % change (increase, due to the hygroscopic nature of formamide) in the volume (found using the spherical cap method) even after 6 h (see Fig S2). Since the sinking time for aqueous drops on a thin lubricating fluid coated surface is about 5-10 min, the environmental exposure does not affect the drop shape during this period.

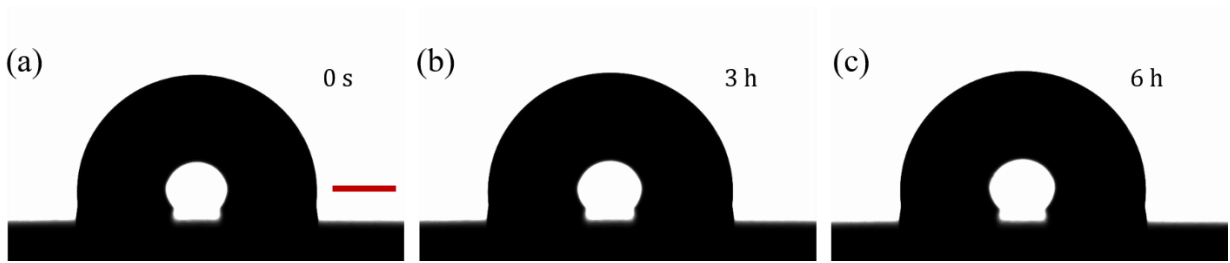

**Figure S2:** Optical images of 80 wt% formamide drop on a hydrophonbic surface after (a) 0 s (b) 3 h and (c) 6 h. The drop volume increased from 2.02 µl (0 s) to 2.16 µl (3 h) and 2.25 µl (6 h) under the lab condition. The scale bar corresponds to 500 µm.

## S3: To find surface and interfacial tension using the pendant drop method:

Various drop-air and drop-oil interfacial tensions were found using the standard pendant drop method with the help of the contact angle goniometer. The optical images of 80 wt % formamide and ethylene glycol pendant drops in air and silicone oil are shown in Fig. S3. Experimentally obtained values are shown in Table S1.

| Drop | Interfacial tension (mN/m) | |
| --- | --- | --- |
| | Drop-air | Drop-oil |
| 80 wt % formamide | 60.7 (0.2) | 26.0 (0.5) |
| Ethylene glycol | 48.3 (0.1) | 16.9 (0.2) |

**Table S1:** Interfacial tensions of various drop liquids measured using the pendant drop method.

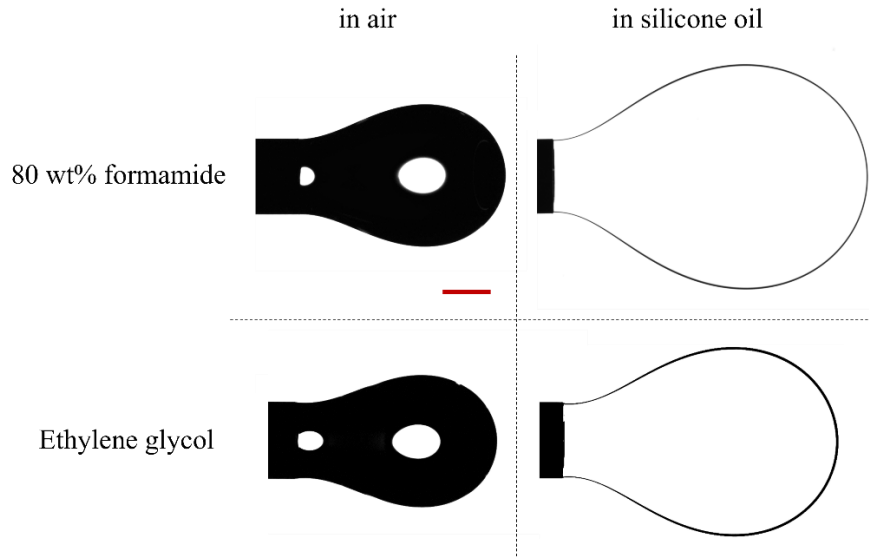

**Figure S3:** Optical images of 80 wt% formamide and ethylene glycol (EG) drops in air and in 350 cSt silicone oil just before detaching from the needle to find the surface/interfacial tensions using the pendant drop method. The scale bar corresponds to 1 mm.

## S4: To find the thickness of spin-coated lubricant films:

Thickness of lubricant films prepared at different spin coating rotation speeds (with fixed spin time and acceleration time as 100 s and 10 s, respectively) was measured using reflected spectrum from Filmetrics F20-EXR, as shown in Fig. S4 (a). The measured reflected spectra is fitted with a calculated model to extract the thickness of the film. Figure S4 (b) shows the thickness of the lubricated films obtained using different rotation speeds in the spin coater. The solid blue line represents the fitted curve, Eq. 3, indicating that the thickness decreases with power law, exponentially with the increasing rotation speed $\omega$.

$$t = t_0 + Ae^{-B\omega} \qquad (3)$$

where $t_0 = 1.45$ μm, $A = 23.94$ μm, and $B = 6.38 \times 10^{-4}$ rpm$^{-1}$.

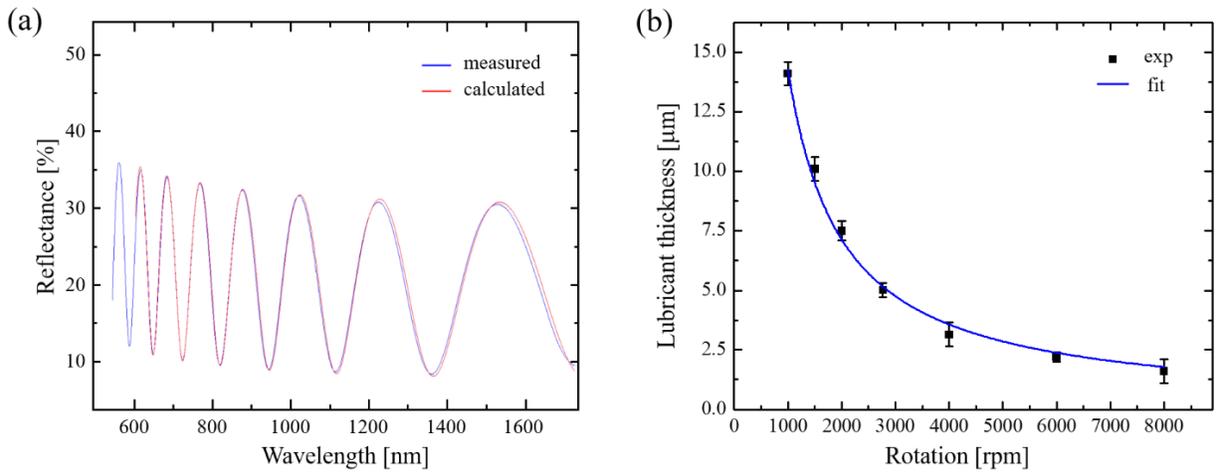

**Figure S4:** (a) Measured (in blue) and calculated (in red) reflectance with wavelength using the Filmetrics-F20-EXR device for a thin silicone oil film prepared at 6000 rpm. (b) Plot showing the variation of silicone oil thickness with the rotation speed in rpm. Solid blue line represents the inverse power-law fit to the experimental data.

## S5: Generalizing the simulation for drops of different liquids:

To check the validity of the simulations for other systems, we used ethylene glycol (EG) and water as the other test drops. The interfacial tension of the water-air and water-silicone oil was taken as 72 mN/m and 39 mN/m, respectively, from the literature [1]. The interfacial tension of the EG-air and EG-silicone oil was found from the pendant drop method (see table S1). Contact angles of EG and water drops on substrate ($\theta_{DS,e}^{O}$) in air and oil was found using the contact angle goniometer as shown in Fig. S5. SE simulations were then carried out for these drops to generate the drop profiles resulting in a good agreement between the simulations (solid red profiles) and the experiments (optical images) as shown in Fig S5.

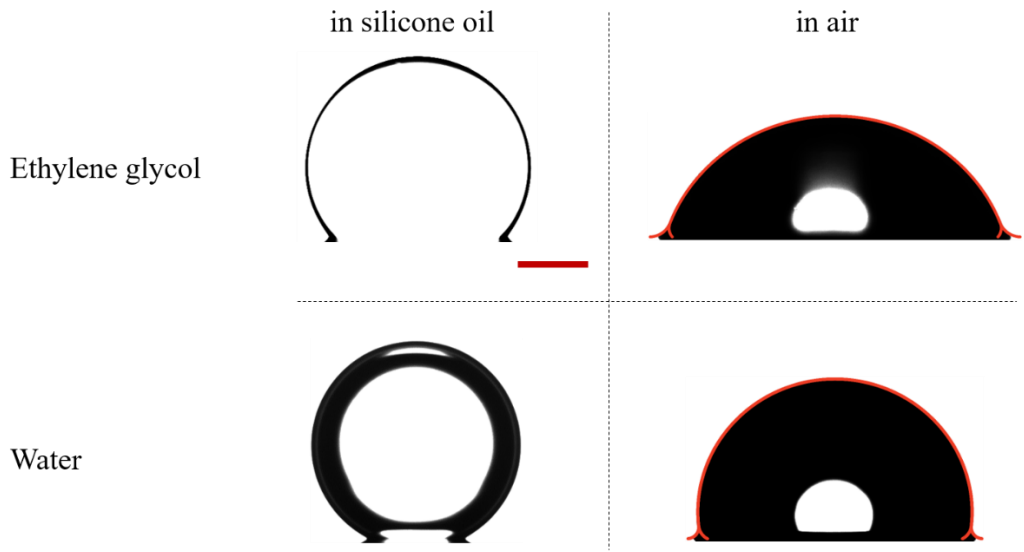

**Figure S5:** (i) Ethylene glycol and (ii) water drops on silicon surfaces with contact angles $\theta_{DS,e}^{O}$ = 130.7 (2.3)° and $\theta_{DS,e}^{O}$ = 153.3 (3.0)° in (a) ambient silicone oil and (b) ambient air with overlapped experimental and surface evolver simulation results. The scale bar corresponds to 500 μm.

**References:**

[1] V. Bergeron, D. Langevin, Monolayer Spreading of Polydimethylsiloxane Oil on Surfactant Solutions, Phys. Rev. Lett. 76(17) (1996) 3152-3155.